\documentclass{IEEEtran}
\usepackage{amsmath,amsthm}
\usepackage{authblk}
\usepackage{algorithmic}
\usepackage{algorithm}
\usepackage{stmaryrd}
\usepackage{pxfonts}
\usepackage{graphicx}
\usepackage{array}
\usepackage{flushend}
\title{Trust Estimation in Peer-to-Peer Network Using BLUE}
\author{Ruchir Gupta, Yatindra Nath Singh, \emph{Senior Member IEEE}}
\affil{Department of Electrical Engineering, IIT, Kanpur}
\affil{\textit {\{rgupta,ynsingh\}@iitk.ac.in}}
\date{16.5.12}

\begin{document}
\maketitle
%\begin{doublespace}
%\markboth{IEEE TRANSACTIONS ON KNOWLEDGE AND DATA ENGINEERING,~Vol.~*, No.~*, *****~****}%
%{Shell \MakeLowercase{\textit{et al.}}: Bare Demo of IEEEtran.cls for Computer Society Journals}
%\IEEEcompsoctitleabstractindextext{%
\begin{abstract}
In peer-to-peer networks, free riding is a major problem. Reputation management systems can be used to overcome this problem. Reputation estimation methods generally do not consider the uncertainties in the inputs.  We propose a reputation estimation method using BLUE (Best Linear Unbiased estimator) estimator that consider uncertainties in the input variables.
\end {abstract}
 %}

\section{Introduction}
\IEEEPARstart{P}{eer}-to-peer systems have attracted considerable attention in recent past as
these systems are more scalable then client server systems. But distributed
nature of peer-to-peer networks brings many challenges for system designers.
These networks are designed keeping in mind that every node is honest and co-operative. It means, if some node takes some resource from the community, it will also facilitate the community.

But nodes are the entities operated by rational human beings. So nodes are expected to behave in
a selfish manner i.e. they try to maximise their utility. This results in their
non co-operative behaviour. This phenomenon is explained by famous Prisoners
Dilemma, in which Nash Equilibrium (NE) is achieved when both prisoners deceive each other \cite{PD}. Similarly in a file sharing network, if nodes are considered as players, their NE happens when none of them share the resources \cite{Trust Based}.
Tendency of nodes to draw resources from the network and not giving any thing in
return is termed as 'Free Riding'. An Experimental
study\cite{Free Riding on Gnutella} on Gnutella network in 2005 also confirmed
this fact by showing that number of free rider nodes is as high as 85\%.

Such type of problems also exist in e-commerce systems like e-bay. In e-commerce portals like e-bay, people sell and buy different things on-line. Buyers and
sellers generally do not know each other. So the possibility of cheating or possibility of providing a product or service of inferior quality always exists. To avoid this, e-bay uses a rating based reputation system. After every transaction, user gives a feedback rating to his counter part and based on these ratings, reputation is decided. This reputation helps users in making decision about transactions \cite{ebay}.

Trust or reputation management systems can also be used in peer to peer networks
to overcome the problem of free riding as well as to ward off some of the
attacks. We have proposed a reputation aggregation method using differential gossip
in \cite{ruchir}.In a peer to peer file sharing network, trust or reputation of a node
represents its co-operative behaviour towards other nodes. A node seeking some
resource from another node measures the ratio of received resource to the
requested resource after every transaction and uses it to update the trust
value. If  a node imparts all the requested resource, it's reputation is
considered as one and if a node always declines sharing of resources,
it has reputation as zero. 

% The advantage with e-bay like systems is that these have a central server which
%keeps all the feedback rating and reputation related data. So, if a user wants to check any other user's reputation, he just asks the central server and receives authentic information. In peer to peer file sharing networks, as there is no central server,  trust has to be estimated and stored by each node in a distributed fashion.

Good method for trust estimation is needed to design a
good reputation management system. Trust can be estimated in a very simple way as the ratio of received to requested resources but this simple method can not
overcome the effect of noise (i.e. uncertainty) in the estimation of trust
value. We are proposing a trust estimation method using BLUE (Best Linear
Unbiased Estimator) \cite{blue}. This method overcomes the noise effects
considerably and requires almost same amount of memory and computation.

Remainder of this paper is organised as follows: Section two discuses the
related work in reputation management while section three describes the system model. In section four the estimator for trust using BLUE is derived. Section six presents the numerical results and section seven concludes the paper.

\section{Related Work}
Different authors  have used different methods for estimation of
trust value of a node. It is also termed as the contribution from the node.
Buragohain \emph{et.al.} \cite{A Game Theoretic: Chiranjeeb} and Wang
\emph{et.al.} \cite{to play or to control} take the ratio of resource
contributed by the node to the ratio of absolute measure of contribution, 
multiplied by the cost of resource. Buragohain \emph{et.al.} does not discuss
the mechnism to measure of contributions of a node by the receiving node. In
\cite{Cooperative Peer Groups in NICE} receiving node computes the trust value
of a node on the basis of received data in the transactions with the sending
node. In \cite{Cooperative Peer Groups in NICE}, quality of transaction is sent
to service provider node in form of a cookie. Duttay \emph{et.al.} \cite{The
Design of A Distributed Rating Scheme for Peer-to-peer Systems} and Papaioannou
\emph{et.al.} \cite{Reputation based policies} suggest that each node should
calculate reputation of other node on the basis of service received from the other nodes depending upon
number of transactions done with those nodes, delay in the transactions and the
download speed. Andrade \emph{et.al.} \cite{Discouraging free riding in a
peer-to-peer CPU-sharing grid} calculates the reputation of a node by taking the
difference of resources received from and provided to the node. In \cite{Limited
Reputation Sharing in P2P
Systems},\cite{eigentrust},\cite{PEERTRUST},\cite{powertrust} and \cite{gossiptrust}, 
a node adjusts the reputation of other node on the basis of quality of transactions
with that node. Eigen-Trust \cite{eigentrust} uses sum of positive and negative
ratings, Peer-Trust \cite {PEERTRUST} normalises the rating on each transaction
whereas Power-Trust \cite{powertrust} uses Bayesian approach to calculate reputation locally.

Mengshu \emph{et.al.} \cite{A trust model of p2p system based on confirmation
theory} takes the ratio of successful transactions to total transactions. PET
\cite {Pet} classifies the service quality into four types (every type provides
some score), then calculates ratio of score obtained by a node to total score it
could have got. After calculating this ratio, it is combined with feedback
obtained from different nodes and final reputation is calculated. Banerjee
\emph{et.al.} \cite{reciprocal} calculates reputation of a node for a particular
type of resource by taking the ratio of number of times this type of resource
was served by that node to the number of times this type of resource was asked from that
node. In Fuzzy-Trust \cite{FuzzyTrust}, \cite{fuzzy approach}, nodes do
\emph{fuzzy inference} on parameters to calculate the trust score locally for
another node 'x' and then aggregate it with trust scores of the node 'x' as
received from other nodes using their weights. Peer can also give rank to the
interacting nodes on the basis of normalized download volume received from them during a fixed period \cite{multi lavel tit for tat}.  The rank of an interacting node is computed using a function that increases when the node provides a resource and decreases when a resource is allocated to the node \cite{Ranking-based Optimal Resource Allocation}. In \cite{A Reputation-based Trust Management}, peer maintains a binary vector of m bits. After a transaction happens one or zero is added at most significant position of the vector, after shifting right all the previously placed bits by one place. This trust vector is considered as a m bit binary number. To compute the trust, value of the m 
bit binary number is divided by $2^{m}$. This ensures that the trust value lies in between 0 and 1. In \cite{R2P}, peer takes the ratio of the total contribution received from a node to the sum of total contribution received from that node and total allocation done by the peer for that node. In \cite{Reputation-based resource allocation}, peer calculates the ratio of the sum of resource it has received from a node to the resource it has requested  to that node for last ten transactions. Moon \emph{et.al.} \cite{Point Based} classifies the services from a node and assigns  different weights to different services, \emph{e.g.}  weight for file transfer is $0.4$, weight for query response is $0.2$ etc.. Peer gives score to the interacting node accordingly.
\section{System Model}
In this paper we are studying a peer-to-peer network. There is no dedicated server in this network. Peers in this network are rational, i.e. they are only interested in their own welfare. They are connected to each other by an access link followed by a back bone link and then again by an access link to the second node.We are assuming that the network is heavily loaded i.e. every peer has
sufficient number of pending download requests, hence these peers are contending for the available transmission capacity. We also assume that every peer is paying the cost of access link as per the use. So, every peer wants to maximises it's download and minimise it's upload so that it can get maximum utility of it's spending and this leads to problem of free riding.

If a node is downloading, some other node has to upload. So the desired condition is, download should be equal to upload for a node. Usually this means that there is no gain. Even in this scenario the node gains due to interaction with others, chance of survival increases. Thus interaction itself is an incentive. A node will usually try to get the content and avoid uploading it as this maximises gain for it. Thus free riding becomes optimal strategy. So a reputation management system need to be enforced to safeguard the interest of every node by controlling free riding.

In reputation management system, every node maintains a reputation table. In this table, the node maintains the reputations of the nodes with whom it has interacted. Whenever it receives a resource from some node, it adjusts the reputation of that node accordingly. When a node asks for the resource from this node, it checks the reputation table and according to the reputation value of requesting node, it allocates resource to that node. This ensures that every node is facilitated from the network as per its contribution to the network and consequently free riding will be discouraged.

For using such a reputation management system node needs to estimate the trust value of the nodes interacting with it. This estimation can be made on the basis of measured trust values  after every transaction. These measurements can be done using any way described in literature. For simulation purpose we have considered trust value observed by node i for node j in transaction k as follows,

\begin{equation}\
  x^{k}_{ij}=\frac {Z^{k}_{ij}}{R^{k}_{ij}}.
\end{equation}

Here $ R^{k}_{ij} $ represents the amount of resources requested by node i from node j
for $k^{th}$ transaction;  $ Z^{k}_{ij} $ represents the amount of resource
received by node i from node j in $k^{th}$ transaction.

In estimation of trust value by such methods few important points are missed out. These points are as follows.
\begin{enumerate}
\item In p2p networks, when a peer asks for some resource, it is not guaranteed that it will get the asked resource. So, generally peer asks for larger amount of resource than needed and when it is offered more resource then its requirement, it refuses the extra offered resource. Also it does not give any credit for this extra offer.
\item Once requesting node decides about the node from which it is going to take data, both of the nodes decide about the rate of data according to their upload and download capacities. But at the some time underlying network may not able to provide this kind of data rate because of congestion in the network at different routers. So, even if service provider node is trying to give data at the committed rate to the requesting node, it can not give and this affects the reputation assignment as reputation is assigned based on the actual data rate.
\item If some requested node has already got too many requests, it will not be able to provide the quality of service that it could have provided with lesser load. Hence the requesting node needs to estimate the reputation of service provider node considering the load and previous transactions with that node.
\end{enumerate}
To resolve above issues, the trust value should be calculated according to the following method:
if node j offered the data against the request of node i and it didn't accept the offer.
\begin{equation}
t_{ij}=\delta,
\end{equation}
Here $\delta$ should be upper bounded in such a way that no node can play game for its benefit. It means if value of $\delta$ will be too high, node will made an offer even if it don't want to serve and on asking for resource it will deny. So node will upper bound it by the ratio of its download capacity to the total requests made by the node.

Whereas if node i accepts the offer
\begin{equation}
t_{ij}=\frac{actual\; service\; rate}{feasible\; service\; rate} \times \frac{willing\; service\; rate}{requested\; service\; rate}
\end{equation}
Here actual service rate is the average rate at which receiver received the data and feasible service rate is the rate at which the TCP Reno algorithm can get the throughput via underlying link with packet loss probability $p$. This rate can be computed using the expression given in \cite{tcpreno}.
\begin{equation}
B(p)\approx \left(\frac{W_{max}}{RTT}, \frac{1}{RTT\sqrt{\frac{2bp}{3}}+T_{0}\cdot min\left(1,3\sqrt{\frac{3bp}{8}}\right) p(1+32p^{2})}\right)
\end{equation}
Here $B$ is feasible service rate as a function of packet loss probability $p$; $W_{max}$ is the maximum window opened by receiver, RTT is the round trip time between two nodes; $T_{0}$ is timeout period and $b$ is number of packets acknowledged by single a acknowledgement.

 Ratio of willing service rate and requested service rate (let's call it $A$) needs to be estimated on the basis of observed samples of the ratio of offered service rate and requested service rate (let's call it $x[n]$, where $n$ is time instant). Let the model for the estimation of $A$ be

\begin{equation}\
\label{model}
  x[n]= A - w[n],
\end{equation}
Here $x[n]$ is the observed value for a node in a particular transaction, $A$ is the exact value and $w[n]$ is the noise in estimating the exact value. Let us assume that $A$ is constant and mean value of $w[k]$ is $W$ and the ratio of $W$ and $A$ is $C$ 
\begin{equation}
C=\begin{cases}
       1-\frac{1}{C_{1}\times C_{2}},& \text{if $C_{1}\times C_{2}$} > 1.\\
			0, & \text{otherwise}.
  \end{cases}
\end{equation}

 Here
\begin{eqnarray}
C_{1}&=&\frac{requests \;made\; by\; the\; node}{download\; capacity \;of
\;the\; node}\\
C_{2}&=&\frac{total\; capacity \;shared\; by \;the\; nodes\; in\; the
\;network}{total\; requests\; made \;by\; nodes\; in\; the\; network}
\end{eqnarray}

Here $C_{1}$ is the ratio of the total request made by that node and download capacity of the node i.e. how much node over requested, and $C_{2}$ is the ratio of total capacity shared by all the nodes  and total requests made in the network by all the nodes.
$C_{2}$ is statistically the average offer against unit request by a node
in the network. Multiplying $C_{2}$ with $C_{1}$ will give us the average
offer to the node against its capacity. Uncertainty have identical distribution for every sample and samples are independent, i.e. the noise samples are i.i.d.. Hence every sample will have same variance. Let us assume this variance be $\sigma$.

\section{Estimation of Trust}
Based on observed values of trust, exact trust value can be estimated using some
estimator. We have used Best Linear Unbiased Estimator (BLUE) \cite{blue} for
this purpose.
Taking expectation of $ x[k] $ in eqn(~\ref{model})
\begin{eqnarray}\
  E[x[k]] &= &E[A] - E[w[k]], \\
  \nonumber & =& A-W.
\end{eqnarray}

so the scaled mean will be
\begin{equation}\
  S[k]= \frac{E[x[k]]}{\Theta}
\end{equation}

where $\Theta$ is the parameter to be estimated i.e. A in this case, so

\begin{eqnarray}
   \nonumber S[k] &=& \frac {A-W}{A} \\
  &=& 1-\frac{W}{A}
\end{eqnarray}

so

\begin{equation}\
  S = \biggl(1-\frac {W}{A}\biggl){\bf 1_{1xN}}
\end{equation}
Here ${\bf 1_{1xN}}$ is a $1\times N$ matrix of 1s i.e. [1 1 1 1 1 1.............1].\\
The covariance matrix
 \begin {equation}\
  \bf  C = \begin{bmatrix}

      \sigma^{2} & 0\\

      0 & \sigma^{2} & 0\\

      & \ddots & \ddots \\

      & & 0 & \sigma^{2}

\end{bmatrix}_{N\times N}
\end{equation}
\begin{equation}\
\bf {C} = \sigma^{2}\bf{I}_{NxN}
\end{equation}
As we know that the BLUE \cite{blue} is

\begin{equation}\
\hat {A}  = \bf{\frac{S^{t}C^{-1}X}{S^{t}C^{-1}S}}
\end{equation}

substituting the values,

\begin{equation}\
\hat {A} = \frac{\biggl(1-\frac{W}{A}\biggl)\bf{1^{t}}\frac{1}{\sigma
^{2}}\bf{I_{NxN}X}}{\biggl(1-\frac{W}{A}\biggl)\bf{1^{t}}\frac{1}{\sigma
^{2}}\bf{I_{NxN}}\biggl(1-\frac{W}{A}\biggl)\bf{1}}
\end{equation}

solving, we get

\begin{eqnarray}\
\hat{A}&=&\frac{\frac{1}{N}\sum\limits^{N}_{k=1}x[k]}{1-\frac{W}{A}}\\
\nonumber &=&\frac{\overline{x[N]}}{1-\frac{W}{A}}
\end{eqnarray}

So, we can see that we need to compute the sample mean of all samples and the
ratio of noise mean and parameter mean.

This formulation have two problems, first it is difficult to compute the sample mean when the number of samples is large and second, if value of $A$ is changing after some time, this estimator doesn't consider that change. So, we can use exponential moving average of samples i.e.
\begin{equation}
\overline {x [N]}=\alpha\cdot x[N] + (1-\alpha)\cdot\overline{ x[N-1]}
\end{equation}
Here $\alpha$  depends on the rate of change in behaviour of nodes. We have evaluated the performance for three different values of $\alpha$ i.e. $0.1$, $0.01$ and $0.001$ whereas $\overline{x[1]}=x[1]$.

Value of $C_{2}$ is estimated regularly on the basis of its download capacity and total requests made. whereas value of $C_{2}$ for complete network is difficult to find. High degree nodes will regularly gather the capacities shared and requests made by the their neighbouring nodes. On the basis of this data high degree nodes will evaluate the value of $C_{2}$. Node will periodically ask neighbouring high degree node(s).

As shown in related work, estimation of trust has been done by a number of ways in literature. We are just considering just one of these ways. But the uncertainties are generally not considered in all of these. So our method can be used for all of them.
\section{Numerical Results}
Performance of algorithm for reputation estimation for peer to peer file sharing is evaluated by simulation as well. 
\subsection{Trust Estimation}
\begin{figure}[!t]
\begin{center}
\includegraphics[width=85mm, height=85mm, keepaspectratio=false]{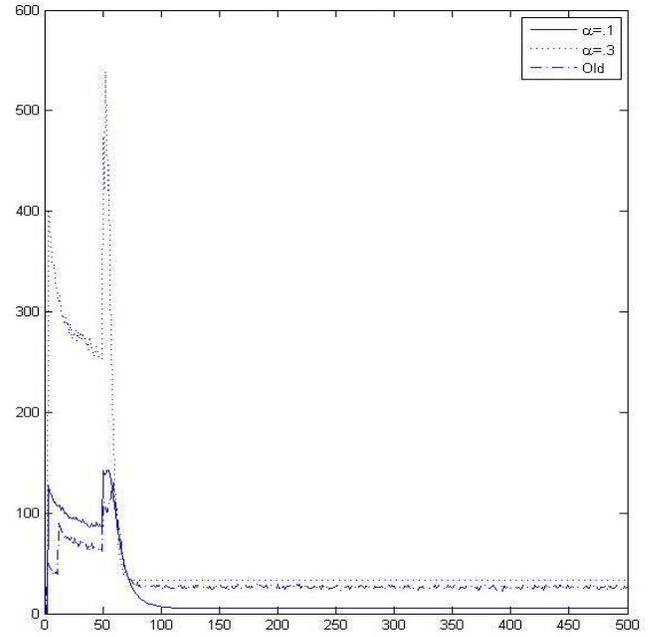}
\caption{Average absolute change in reputation for homogeneous network}
\label{fig:changerephomo}
\end{center}
\end{figure}

\subsection{Trust Estimation}
\begin{figure}[!t]
\begin{center}
\includegraphics[width=85mm, height=85mm, keepaspectratio=false]{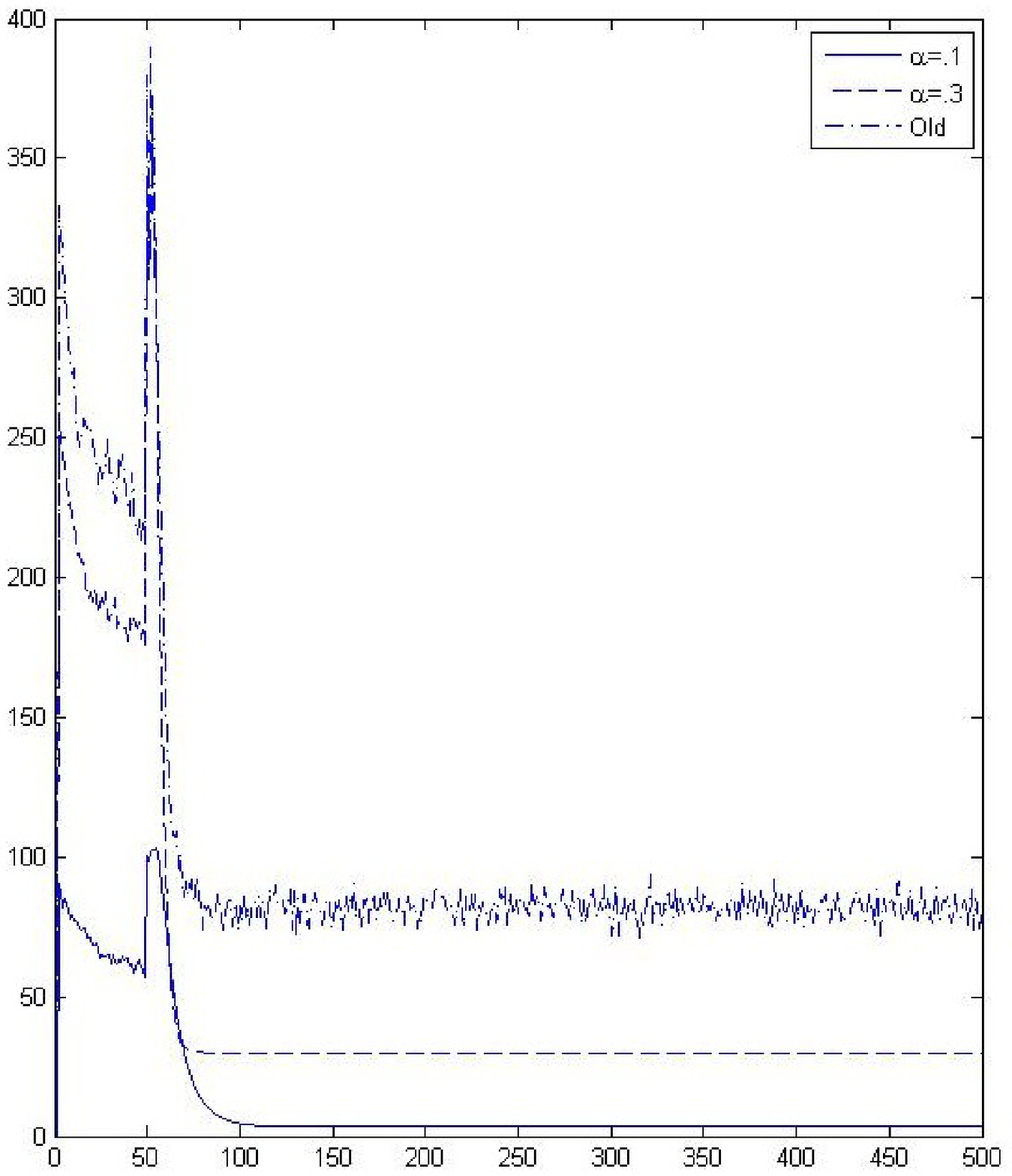}
\caption{Average absolute change in reputation for heterogeneous network}
\label{fig:changerephtr}
\end{center}
\end{figure}

\begin{figure}[!t]
\begin{center}
\includegraphics[width=85mm, height=85mm, keepaspectratio=false]{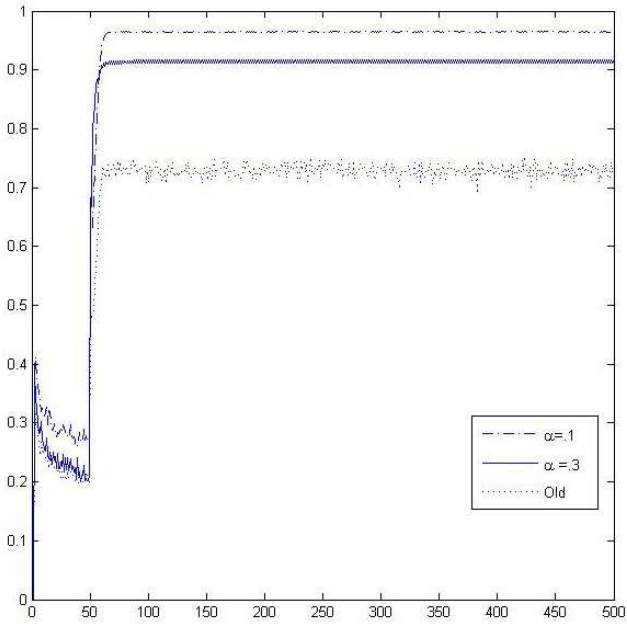}
\caption{Network utilisation by a homogeneous network of 200 nodes}
\label{fig:networkutilhomo}
\end{center}
\end{figure}

\begin{figure}[!t]
\begin{center}
\includegraphics[width=85mm, height=85mm, keepaspectratio=false]{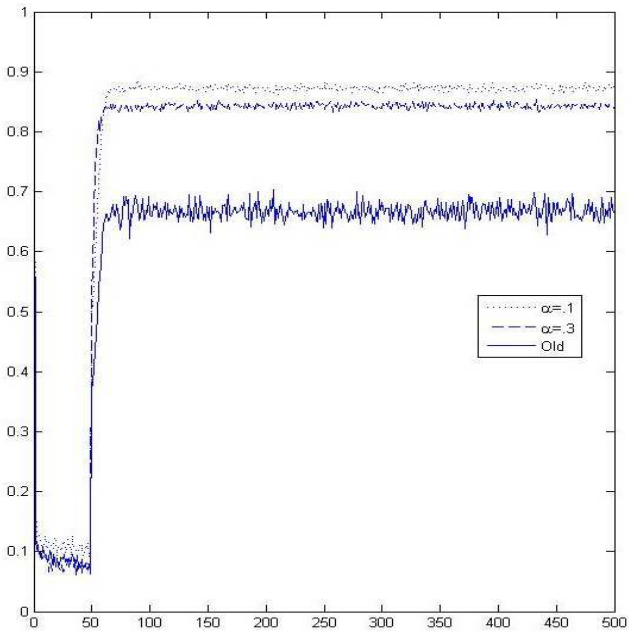}
\caption{Network utilisation by a heterogeneous network of 200 nodes}
\label{fig:networkutilhtr}
\end{center}
\end{figure}

Performance of estimation method proposed in this paper has been evaluated for $200$ node network. Time is considered as slotted. Every slot is termed as an iteration. At the start of every slot every node queries for some resource and after getting reply it asks for resource from replying nodes. Node asks for resource as per its download capacity from the nodes with resource. Requested nodes allocate their bandwidth as per their reputation table in a probabilistic manner. Number of requests that a node will serve is fixed. For homogeneous network its same for every node. Where as its different in heterogeneous network.  Finally requesting node and requested nodes hand shake and transect. At the end of slot each node updates its reputation table as per the quality of transaction. First $50$ iterations have been taken as acquittance period i.e.  node will allocate their bandwidth without referring the reputation table. Value of $\delta$ has been assumed as $0.3$. Reputation is measured by node $i$ for node $j$ in transaction $k$ has been defined as,
\begin{equation}\
  R^{k}_{ij}=\frac {Alo^{k}_{ij}}{Req^{k}_{ij}}.
\end{equation}

Here $ Req^{k}_{ij} $ represents the amount of resources requested by node $i$ from node $j$
for $k^{th}$ transaction;  $ Alo^{k}_{ij} $ represents the amount of resource
received by node $i$ from node $j$ in $k^{th}$ transaction.

We have plotted the average absolute change in reputation estimation with increasing number of iterations up to $500$. Here, absolute change in reputation ($\Delta R$)is
\begin{equation}
\Delta R(itr)=\sum\limits_{i,j}|R_{i,j,itr}-R_{i,j,itr-1}|
\end{equation}
$R_{i,j,itr}$ is the reputation of $j$ for $i$ in ith iteration, $d_{i}$ is the degree of ith node, $N$ is the total number of nodes and $itr$ is the number of iteration under consideration. 

In figure figure ~\ref{fig:changerephomo} and ~\ref{fig:changerephtr} $\Delta R$ is plotted by taking the average of last ten measurements \cite{Reputation-based resource allocation} and proposed reputation estimation method using these measurements for homogeneous and in figure ~\ref{fig:changerephtr} for heterogeneous networks for 200 nodes with $\alpha$= 0.1 and 0.3. Here by homogeneous network we mean a network where every node has same download capacities and is ready to serve same number of nodes. Whereas by heterogeneous network we mean that nodes have different download capacities and are ready to serve different number nodes.

This is evident in both figure ~\ref{fig:changerephomo} and ~\ref{fig:changerephtr} that using estimator the change in reputation is less. This implies that by use of proposed estimator reputation can be estimated more accurately. 

In figure~\ref{fig:networkutilhomo} and in figure~\ref{fig:networkutilhtr} the utilisation level of shared resources in the network is plotted for  homogeneous and heterogeneous networks of 200 nodes for $\alpha$ $0.1$ and $0.3$. 

\section{Conclusion}
In peer-to-peer networks, free riding is a major problem that can be overcome by using reputation management system. A reputation management system includes two processes, first estimation of reputation and second aggregation of reputation. In this paper we have proposed an estimation technique using BLUE. 

The proposed trust estimator considers uncertainties in the trust estimation.  The average absolute change in trust values in the trust table of nodes is considerably less then the older techniques. It implies that reputation can be estimated more accurately using this estimator. The better estimation of reputation will lead to a better utilisation of resource in the system as evident. 


\begin{thebibliography}{1}

\bibitem{PD}
M.J. Osborne,"An Introduction to Game Theory". \emph{Oxford University Press}, Oxford (2002)

\bibitem{Trust Based}
W. Wang and B. Li., "Trust Based Incentive in P2P Network". \emph{Proc. Int.
IEEE Conf. on E-Commerce Technology for Dynamic E-Business (CEC-East)}, pages
302305, 2004

\bibitem{Free Riding on Gnutella}
D. Hughes, G. Coulson, and J. Walkerdine, Free Riding on Gnutella Revisited:
The Bell Tolls, \emph{IEEE Distributed Systems Online}, vol. 6, no. 6, June
2005.

\bibitem{ebay}
http://www.ebay.in/

\bibitem{A Game Theoretic: Chiranjeeb}
Chiranjeeb Buragohain, Divyakant Agrawal and Subhash Suri, "A Game
Theoretic Framework for Incentives in P2P Systems", Proceedings of the \emph{3rd
International Conference on Peer-to-Peer Computing}, p.48, September 01-03, 2003.

\bibitem{to play or to control}
Weihong Wang, Baochun Li, "To Play or to Control: A Game-based
Control-theoretic Approach to Peer-to-Peer Incentive Engineering", in the
Proceedings of the \emph{Eleventh IEEE International Workshop on Quality of Service (IWQoS 2003)}, also \emph{Lecture Notes in Computer Science, ACM Springer-Verlag}, vol. 2707, pp. 174-192, Monterey, CA, June 2-4, 2003.

\bibitem{Cooperative Peer Groups in NICE}
Seungjoon Lee, Rob Sherwood and Bobby Bhattacharjee, "Cooperative Peer
Groups in NICE", \emph{Journal of Computer Networks (COMNET) Special Issue on Trust and Reputation in Peer-to-Peer Systems}, 2005.

\bibitem {The Design of A Distributed Rating Scheme for Peer-to-peer Systems}
Debojyoti Duttay, Ashish Goelz, Ramesh Govindany and Hui Zhangy, "The
Design of A Distributed Rating Scheme for Peer-to-peer Systems", Proceedings of
\emph{the Workshop on the Economics of Peer-to-Peer Systems}, 2003.

\bibitem{ruchir}
Ruchir Gupta, Y. N. Singh," Reputation Aggregation in Peer-to-Peer Network Using Differential Gossip
Algorithm", arXiv:1210.4301 [cs.NI].

\bibitem{Reputation based policies}
Thanasis G. Papaioannou , George D. Stamoulis,"eputation-based policies
that provide the right incentives in peer-to-peer environments", \emph{Computer
Networks: The International Journal of Computer and Telecommunications
Networking}, v.50 n.4, p.563-578, 15 March 2006

\bibitem{Discouraging free riding in a peer-to-peer CPU-sharing grid}
N. Andrade, F. Brasileiro, W. Cirne, M. Mowbray, "Discouraging free riding in a
peer-to-peer CPU-sharing grid", \emph{HPDC 04: Proceedings of the 13th IEEE
International Symposium on High Performance Distributed Computing},
IEEE Computer Society, Washington, DC, USA (2004), pp. 129137.

\bibitem{Limited Reputation Sharing in P2P Systems}
S. Marti and H. Garcia-Molina, "Limited Reputation Sharing in P2P
Systems", Proc. \emph{Fifth ACM Conf. Electronic Commerce}, May 2004.

\bibitem{eigentrust}
S. Kamvar, M. Schlosser, and H. Garcia-Molina, "The Eigentrust Algorithm
for Reputation Management in P2P Networks", Proc. \emph{12th Intl World Wide Web Conf. (WWW 03)}, May 2003.

\bibitem{PEERTRUST}
L. Xiong and L. Liu, "PeerTrust: Supporting Reputation-Based Trust for
Peer-to-Peer Electronic Communities", \emph{IEEE Trans. Knowledge and Data Eng.}, vol. 16, no. 7, pp. 843-857, 2004.

\bibitem{FuzzyTrust}
S. Song, K. Hwang, R. Zhou and Y.K. Kwok, "Trusted P2P Transactions with
Fuzzy Reputation Aggregation", \emph{IEEE Internet Computing} pp. 18-28, Nov./Dec.
2005.

\bibitem{Pet}
Z. Liang, W. Shi, "PET: A personalized trust model with reputation and risk
evaluation for P2P resource sharing", in \emph{HICSS-38}, Hilton Waikoloa Village Big Island, Hawaii, January 2005

\bibitem{powertrust}
R. Zhou and K. Hwang, emph, "PowerTrust: A Robust and Scalable Reputation System
for Trusted P2P Computing", \emph{IEEE Trans. Parallel and Distributed Systems}, vol. 18, no. 4, pp. 460-473, Apr. 2007.

\bibitem{gossiptrust}
Runfang Zhou , Kai Hwang , Min Cai, "GossipTrust for Fast Reputation
Aggregation in Peer-to-Peer Networks", \emph{IEEE Transactions on Knowledge and Data Engineering}, v.20 n.9, p.1282-1295, September 2008

\bibitem{A trust model of p2p system based on confirmation theory}
Hou Mengshu , Lu Xianliang , Zhou Xu , Zhan Chuan, "A trust model of p2p
system based on confirmation theory", \emph{ACM SIGOPS Operating Systems Review}, v.39 n.1, p.56-62, January 2005.

\bibitem{reciprocal}
D. Banerjee, S. Saha, S. Sen, and P. Dasgupta, "Reciprocal resource
sharing in p2p environments", In \emph{4th Int. Joint Conference on Autonomous Agents and Multi Agent Systems}, 2005.

\bibitem{fuzzy approach}
Ansuman Mahapatra, Nachiketa Tarasia, Anuja Ajay, Soumya Ray, "A Fuzzy
Approach for Reputation Management in Bittorrent P2P Network", \emph{3rd International Conference on Electronics Computer Technology}, ICECT 2011,IEEE, 8-10 Apr 2011.

\bibitem{multi lavel tit for tat}
LIAN, Q., PENG, Y., YANG, M., ZHANG, Z., DAI, Y., AND LI, X., "Robust
incentives via multi-level tit-for-tat". In Proc. of \emph{IPTPS (Santa Barbara, CA}, Feb. 2006).

\bibitem{Ranking-based Optimal Resource Allocation}
Yan, Y., El-Atawy, A., and Al-Shaer, E. 2007. "Ranking-based Optimal
Resource Allocation in Peer-to-Peer Networks". In Proceedings of the \emph{26th Annual IEEE International Conference on Computer Communications (INFOCOM)}.

\bibitem{A Reputation-based Trust Management}
A. Selcuk, E. Uzun, M.R. Pariente, "A reputation-based trust management
system for P2P networks", in \emph{Fourth Intl Workshop on Global and Peer-to-Peer Computing}, April 2021, 2004.

\bibitem{R2P}
Y. Xia, G. Song, Y. Zheng, and M. Zhu. "R2p: A peer-to-peer transfer system
based on role and reputation". \emph{First International Workshop on Knowledge Discovery and Data Mining}, pages 136141, Jan. 2008.
\bibitem{Reputation-based resource allocation}
A. Satsiou, L. Tassiulas, "Reputation-based resource allocation in P2P
systems of rational users", \emph{IEEE Trans. Parallel Distrib. Syst.}, 21 (4) (2010),
pp. 466479.

\bibitem{Point Based}
Jongbae Moon, Yongyun Cho, "A point-based inventive system to prevent
free-riding on P2P network environments", Proceeding \emph{ICCSA'11 Proceedings of the 2011 international conference on Computational science and its applications} -
Volume Part IV Pages 462-471.


\bibitem{tcpreno}
J.D.Padhye, V.Firoiu, D.F. Towsley, and J.F.Kurose, "Modeling TCP throughput: A simple Model and its empirical validation", In proceedings of \emph{ACM SIGCOMM}, pp.303-314, Vancouver, Canada, September 1998.
\bibitem{blue}
S.M.Kay, "Fundamentals of Statistical Signal Processing:Estimation
Theory", \emph{Prentice Hall}, 1993.

\end{thebibliography}
\end{document}